\documentclass[11pt]{article}
\usepackage{amsfonts}
\newcommand{\be}{\begin{equation}}
\newcommand{\ee}{\end{equation}}
\newcommand{\bea}{\begin{eqnarray}}
\newcommand{\eea}{\end{eqnarray}}
\newcommand{\rb}{\right]}
\newcommand{\lb}{\left[}
\renewcommand{\(}{\left(}
\renewcommand{\)}{\right)}
\newcommand{\ao}{a_n^{(1)}}
\newcommand{\ato}{a_2^{(1)}}
\newcommand{\uq}[1]{$U_q(#1)$}
\newcommand{\cR}{\check{R}}
\newcommand{\tK}{\tilde{K}}
\newcommand{\nn}{\nonumber}
\newcommand{\ol}{\overline}
\newcommand{\ul}{\underline}
\newcommand{\noi}{\noindent}
\newcommand{\hs}{\hspace}
\newcommand{\vs}{\vspace}
\newcommand{\lra}{\longrightarrow}
\newcommand{\ot}{\otimes}
\advance\textheight by \topskip
\makeatletter

\@addtoreset{equation}{section}
\def\section{\@startsection {section}{1}{\z@}{-8.5ex plus -1ex minus
 -.2ex}{3.3ex plus .2ex}{\large\bf\centering}}
\def\subsection{\@startsection{subsection}{2}{\z@}{-3.25ex plus
 -1ex minus -.2ex}{1.5ex plus .2ex}{\bf}}
\def\subsubsection{\@startsection{subsubsection}{3}{\z@}{-3.25ex plus%
 -1ex minus -.2ex}{1.5ex plus .2ex}{\sl}}

\begin{document}

\newpage
\begin{titlepage}
\begin{flushright}
DTP--99/63 \\
hep-th/9911178 \\
October 1999
\end{flushright}
\vspace{1cm}
\begin{center}
{\Large {\bf New non-diagonal solutions to the\\  $\ao$ boundary
Yang-Baxter equation}}\\
\vspace{1.5cm}
{\large G.\ M.\ Gandenberger}\footnote{\noi E-mail:
georgg@globalnet.co.uk}\\
\vs{1cm}
{\em Department of Mathematical Sciences\\
Durham University\\
Durham DH1 3LE, U.K.}\\
\vspace{2cm}
{\bf{ABSTRACT}}
\end{center}
\begin{quote}
Extending previous work on $\ato$, we present a set of reflection matrices,
which are explicit solutions to the $\ao$ boundary Yang-Baxter
equation. Unlike solutions found previously
these are multiplet-changing $K$-matrices, and could therefore be used
as soliton reflection matrices for affine Toda field theories on the
half-line.
\end{quote}

\vfill

\end{titlepage}

\section{Introduction}

In the study of two-dimensional integrable models on the whole line an
important role
is being played by the Yang-Baxter-equation (YBE), which arises from
the factorisability of the $S$-matrix. The YBE has been studied for a
long time and a large number of solutions,
the so-called $R$-matrices, are known.
The boundary Yang-Baxter equation (BYBE) is the analogue of the YBE in
two-dimensional models on a half-line, i.e.\ models with one
reflecting boundary. Factorisability and integrability on a half-line
imply a highly non-trivial relation between $R$-matrices and
reflection matrices, which are the so-called $K$-matrices.
Much less work has been done on the BYBE and only a small number of
solutions for some specific cases have been found so far. (For more
details on the BYBE and its connection to integrable models on a
half-line see \cite{ghosh94}, \cite{gande98} and references therein.)

Here we are interested in BYBEs related to trigonometric
$R$-matrices. These are intertwining maps of the form
\[
\cR_{a,b}(x): V_a \ot V_b \lra V_b \ot V_a\;,
\]
in which $x$ is a spectral parameter and the $V$'s are the
representation spaces of the fundamental representations of some
quantized universal enveloping algebra \uq{\hat g} of an
affine Lie algebra $\hat g$. These $R$-matrices satisfy the YBE in the form
\[
\cR_{b,c}(x)\ot I_a\,.\, I_b\ot
\cR_{a,c}(xy)\,.\, \cR_{a,b}(y)\ot I_c = I_c\ot
\cR_{a,b}(y)\,.\, \cR_{a,c}(xy)\ot I_b \,.\, I_a\ot
\cR_{b,c}(x)\;,
\]
in which $I_a$ denotes the identity on $V_a$, such that both sides
of the equation map $V_a\ot V_b \ot V_c$ into $V_c\ot V_b \ot
V_a$. (Note that the indices $a,b,c$ denote the type of representation.)

Here we are only considering the case of $\hat g = \ao$, for which
two different types of the BYBE exist. These two types are
distinguished by whether the $K$-matrices map a representation space
$V_a$ into itself or into the conjugate space $V_{n+1-a}$. These two
cases could be related to integrable models in which the particles either
remain in the same multiplet or transform into a particle in the charge
conjugate multiplet after reflection from the boundary.
The non-multiplet changing BYBE for the lowest $R$-matrices can be
written in the following form:
\be
I_1 \ot \tK(y)\,.\,\cR_{1,1}(xy)\,.\,I_1 \ot
\tK(x)\,.\, \cR_{1,1}(\frac xy) =
\cR_{1,1}(\frac xy) \,.\,I_{1} \ot \tK(x) \,.\,
\cR_{1,1}(xy) \,.\,I_1 \ot \tK(y)\;, \label{nBYBE}
\ee
in which the reflection matrices are maps
\[
\tK(x) : V_1 \lra V_1\;.
\]
Diagonal solutions to equation (\ref{nBYBE}) were found
some years ago in \cite{deveg93} and they have been used in connection
with spin chains (see for instance \cite{doiko98}).

However, more recently it has been discovered that in order to describe
the reflection of solitons in imaginary coupled $\ao$ affine Toda
field theory a different type of $K$-matrices is needed. In
\cite{gande98} it was realised that affine Toda solitons reflect
into antisolitons, which are the states in the charge conjugate
multiplet. This property has also been confirmed in a  semiclassical
study of affine Toda theories on a half-line in \cite{deliu98}.
Therefore, we require $K$-matrices of the general form
\be
K(x): V_1 \lra V_n\;,
\ee
which now have to be solutions to the following multiplet changing BYBE:
\be
I_n \ot K(y)\,.\,\cR_{1,n}(xy)\,.\,I_1 \ot
K(x)\,.\, \cR_{1,1}(\frac xy) =
\cR_{n,n}(\frac xy) \,.\,I_{n} \ot K(x) \,.\,
\cR_{1,n}(xy) \,.\,I_1 \ot K(y)\;.   \label{BYBE1}
\ee
Solutions to this equation for the case of $\hat g = \ato$ were first
found in \cite{gande98} and the solutions for the general case of
$\hat g = \ao$ are the subject of this letter.

\section{The $\ao$ R-matrix}

In order to solve the BYBE (\ref{BYBE1}) we first need an explicit
expression of the $R$-matrix corresponding to the first fundamental
representation of \uq{\ao}.
These $R$-matrices were originally given by Jimbo in
\cite{jimbo86}. We use a slightly modified notation
here\footnote{Comparing this
expression with those in \cite{jimbo86}, we have changed $x \to
x^{2h}$ and $k \to -q^4$. Note also that this is the intertwining
$R$-matrix which is related to Jimbo's $R$-matrix by $\cR = PR$, in
which $P$ is the permutation matrix.}:
\bea
\cR_{1,1}(x) &=& (x^{-h} q^4 - x^h q^{-4}) \sum_{i=1}^n E_{ii}\otimes
E_{ii} + (x^h - x^{-h}) \sum_{i\neq j} E_{ij}\otimes
E_{ji}\nn \\
&&+ (q^4 - q^{-4})\( \sum_{i<j}x^{\nu(i,j)}\,E_{ii}\otimes E_{jj} +
\sum_{i>j} x^{\nu(i,j)}\, E_{ii}\otimes E_{jj}\) \;,
\eea
in which the $E_{i,j}$ are $n\times n $ matrices with the only entry
being a $1$ in the $ith$ row and $jth$ column.  Alternatively,
we can write this $R$-matrix as a $n^2\times n^2$-matrix
$\lb \cR_{1,1}(x)\rb _{i,j}^{k,l}$. For the sake of convenience we
omit the
multiplet labelling from now on and write $\lb \cR_{1,1}(x)\rb
_{i,j}^{k,l} \equiv \cR_{i,j}^{k,l}(x)$. All the non-zero elements of
this matrix are then given by
\bea
\cR_{i,i}^{i,i}(x) &=& x^{-h} q^4 - x^h q^{-4}\;, \nn \\
\cR_{i,j}^{j,i}(x)  &=& x^h - x^{-h} \;, \nn \\
\cR_{i,j}^{i,j}(x) &=& (q^4 - q^{-4})x^{\nu(i,j)}\;,\hs{1cm}
(\mbox{for } i,j = 1,...,n\;\;\;\mbox{and}\;\;\; i \neq j)
\label{Relements}
\eea
in which $h= n+1$ is the Coxeter number of $\ao$, and $\nu(i,j)$ is
determined by the gradation of the $R$-matrices. (For more details
about $R$-matrices and their gradations see for instance
\cite{gande96}, \cite{deliu95}.)
For the two gradations we are interested in, $\nu(i,j)$ is given as
follows:
\bea
\mbox{\em homogeneous gradation:} \hs{2cm} \nu(i,j) &=& \left\{
\begin{array}{ll} -h \hs{2.5cm} & (i<j) \\
h & (i>j) \end{array} \right. \;, \\  \nn \\
\mbox{\em principal gradation:} \hs{2cm} \nu(i,j) &=&
\left\{\begin{array}{ll} 2(i-j)+h \hs{1cm} & (i<j) \\
2(i-j)-h & (i>j) \end{array} \right. \;.
\eea
An important aspect of these $R$-matrices is the fact that they are
crossing symmetric, and one can therefore choose the bases of the
representation spaces such that the crossed $R$-matrices are given as
\bea
\cR_{j,\ol l}^{\ol i, k}(x) &=& \cR_{i,j}^{k,l} (q^2 x^{-1})
\;, \nn \\
\cR_{\ol l,\ol k}^{\ol j,\ol i}(x) &=& \cR_{i,j}^{k,l} (x)
\;, \nn \\
\cR_{\ol k,i}^{l,\ol j}(x) &=& \cR_{i,j}^{k,l}(q^2 x^{-1})
\;. \label{crosselements}
\eea
Here we have used a short hand notation in which barred indices
denote the labels in $V_n$ and unbarred indices those in $V_1$, which means
for instance
\[
\cR_{i,\ol j}^{\ol k,l}(x) \equiv \lb \cR_{1,n}(x)\rb_{i,j}^{k,l}\;:
V_1 \otimes V_n \lra V_n \otimes V_1\;.
\]

\section{The $K$-matrices}

Since the dimension of the spaces $V_1$ and $V_n$ are both equal to $n$,
we can write the $K$-matrices as $n\times n$ matrices $K_i^{\ol j}(x)$.
Using this explicit matrix form we can rewrite the BYBE (\ref{BYBE1})
into the following form:
\be
K_j^{\ol k}(y)\, \cR_{i,\ol k}^{\ol l,m}(xy)\,
K_m^{\ol n}(x)\, \cR_{\ol l,\ol n}^{\ol p,\ol r}(\frac xy) =
\cR_{i,j}^{k,l}(\frac xy)\, K_l^{\ol m}(x)\, \cR_{k,\ol m}^{\ol
p,n}(xy)\, K_n^{\ol r}(y)\;, \label{BYBE2}
\ee
in which summation over repeated indices is implied.

In analogy to the $\ato$ case there is one very simple solution, which
is a diagonal matrix
\be
K_1^{(d)}(x) = \left( \begin{array}{cccccc}
1 & 0 & \dots & & & \\
0 & 1 &  & & & \\
\vdots & & \ddots & & & \\
& & & & & \\
& & & & & 1 \end{array} \right)\;.
\end{equation}
It is straightforward to show that this is indeed a solution to
equation (\ref{BYBE2}). In fact, it is not even necessary to know the
exact form of the $R$-matrix elements, because the diagonal matrix
$K_1^{(d)}$ solves the BYBE corresponding to any $R$-matrix which
satisfies the crossing relations (\ref{crosselements}).
If this $K$-matrix was used as a reflection matrix it would mean that
the states in the first multiplet are reflected into their charge
conjugate partners in the $n$th multiplet. In a recent paper
\cite{deliu99} this solution was used as a starting
point in the construction of the reflection factors for $\ao$
affine Toda field theory.

However, the main result of this letter is the fact that we have also
found two explicit non--diagonal solutions to equation (\ref{BYBE2}),
for both the homogeneous and the principal gradation. These two
solutions are only distinguished by some signs and can be written in
the following form:

\bea
K_i^{\ol j}(x) = \left\{ \begin{array}{cl}
\frac{x^h q^{-h+2} - x^{-h}q^{h-2}} {q^2-q^{-2}}\;, \hs{10pt} &
\mbox{if} \hs{10pt} i=\ol j\;, \\   \\
\(\frac{x}{q}\)^{\nu(i,\ol j)}\;, & \mbox{if} \hs{10pt} i  \neq \ol j \;,
\end{array} \right. \label{Ksol1}
\eea
and
\bea
K_i^{\ol j}(x) = \left\{ \begin{array}{cl}
\frac{x^h q^{-h+2} + x^{-h}q^{h-2}} {q^2-q^{-2}}\;, \hs{10pt} &
\mbox{if} \hs{10pt} i=\ol j\;, \\   \\
\mbox{sign}(\ol j - i) \(\frac{x}{q}\)^{\nu(i,\ol j)}\;, & \mbox{if}
\hs{10pt} i
\neq \ol j \;.
\end{array} \right. \label{Ksol2}
\eea

For the case of $n=2$ these solutions are the same as those found in
\cite{gande98}. Unfortunately, we do not know any elegant way to prove
that these expression do satisfy the BYBE, but they have been checked
in detail using algebraic software. Some details of this check are
provided in the appendix.

In both solutions (\ref{Ksol1}) and (\ref{Ksol2}) there still remains
a $n$-fold freedom, namely these K-matrices remain solutions to the
same BYBE, if we multiply them from the left and from the right by a
diagonal matrix
\be
{\cal E} = \left( \begin{array}{cccccc}
E_1 & 0 & \dots & & & \\
0 & E_2 &  & & & \\
\vdots & & \ddots & & & \\
& & & & & \\
& & & & & E_n \end{array} \right)\;,
\ee
in which $E_i = E_i(q)$ can be arbitrary functions of the deformation
parameter $q$ (but not of $x$). In other words, the BYBE appears to be
invariant under transformations of the form
\be
K_i^{\ol j}(x) \lra E_i(q)\,E_{\ol j}(q)\,K_i^{\ol j}(x)\;.
\ee
If these
$K$-matrices are to be used as reflection matrices for
an integrable model, then the parameters $E_i$ could be related to
different integrable boundary conditions. However, as it was shown in
\cite{gande98},
the additional conditions imposed on the reflection matrices by
boundary unitarity and boundary crossing, can restrict significantly the
possible number of free boundary parameters.

In addition, the $K$-matrices are only determined up to an overall
scalar factor ${\cal A}(x)$. In order to use these solutions as
reflection matrices for the reflection of solitons in $\ao$ affine
Toda field theories, it would be necessary to find explicit
expressions for the overall scalar factor, which is determined by
the boundary-unitarity and boundary bootstrap conditions.

Finally, note that we have only provided solutions related to the first
fundamental representation of \uq{\ao}. In general, there are of course
$n$ possible sets of $K$-matrices $K^{(a)}: V_a \lra V_{n+1-a}$ (for
$a=1,\dots,n)$. These higher $K$-matrices can in principle be
constructed from the solutions (\ref{Ksol1}) and (\ref{Ksol2}) by
using the bootstrap equations. This, however, goes beyond the scope of
this letter. However, we do hope that the explicit solutions found
here will shed some light on the general structure of $K$-matrices and
may help in the construction of $K$-matrices for other algebras.

\vs{1cm}

\noi {\bf \ul{Acknowledgements:}}\\
I would like to thank Peter Bowcock, Ed Corrigan, Gustav Delius and
Patrick Dorey for discussions. This work was supported by an
EPSRC research grant no.\ $GR/K\, 79437\,$ as well as a TMR Network
grant no.\ FMRX-CT-960012.

\vs {1cm}

\appendix
\section{A sketch of proof}

It would be desirable to have a proof in terms of quantum algebra
properties, but all we can do at this stage is to check the above
solutions using some algebraic software such as MapleV. Fortunately,
however, due to the particular form of the $\ao$ $R$-matrices this
check can be performed for general $n$, rather than just for some
particular example. The details of this proof itself are not very
illuminating and we therefore just mention the main idea.

In general, in order to prove equation (\ref{BYBE2}), we would have to
check $n^4$ different equations. However,
because of the fact that most of the entries in the $R$-matrix
(\ref{Relements}) are equal to zero, this number can be decreased
substantially and we can perform the check for general $i,j,k,l = 1,
\dots, n$.
We will demonstrate this for one example in detail.

Consider equation
(\ref{BYBE2}) for the case $i=j$ and $\ol p \neq \ol r$, and also
$i \neq \ol p$ and  $i \neq \ol r$.
Let us first examine the last term on the left hand side of
(\ref{BYBE2}). Because of the fact that $\ol p \neq \ol r$ there are
only two possible non-zero terms, namely $\cR_{\ol p,\ol r}^{\ol p,\ol
r}(\frac xy)$  and $\cR_{\ol r,\ol p}^{\ol p,\ol r}(\frac xy)$.
Thus, the left hand side of the equation reduces to
\bea
\lefteqn{\sum_{\ol k, m} K_i^{\ol k}(y)\, \cR_{i,\ol k}^{\ol p,m}(xy)\,
K_m^{\ol r}(x)\, \cR_{\ol p,\ol r}^{\ol p,\ol r}(\frac xy) +
 K_i^{\ol k}(y)\, \cR_{i,\ol k}^{\ol r,m}(xy)\,
K_m^{\ol p}(x)\, \cR_{\ol r,\ol p}^{\ol p,\ol r}(\frac xy) =} \hs{0.5cm}
\nn \\
&=& \left[ K_i^{\ol i}(y)\, \cR_{i,\ol i}^{\ol p,p}(xy)\,
K_p^{\ol r}(x)\,  +
K_i^{\ol p}(y)\, \cR_{i,\ol p}^{\ol p,i}(xy)\,
K_i^{\ol r}(x)\,\right] \cR_{\ol p,\ol r}^{\ol p,\ol r}(\frac xy) \nn \\
&& \hs{0.5cm} + \left[ K_i^{\ol i}(y)\, \cR_{i,\ol i}^{\ol r,r}(xy)\,
K_r^{\ol p}(x)\, +
K_i^{\ol r}(y)\, \cR_{i,\ol r}^{\ol r,i}(xy)\,
K_i^{\ol p}(x)\,\right] \cR_{\ol r,\ol p}^{\ol p,\ol r}(\frac xy) \nn
\\ &=&
 \(\frac xq \)^{2i-2p} \(\frac yq \)^{2i-2r} \left[ q^4 \(\frac
xy\)^{-h}  - q^{-4} \(\frac xy\)^h \right]\,
\left[ \(\frac {q}{x}\)^{2h}  - \(\frac
{q}{y}\)^{-2h} + (q^8-1) (x^hq^{-h-2} - x^{-h}q^{h+2}) \right] \nn \\
&=&
\(\frac xq \)^{2p-2i} \(\frac yq \)^{2r-2i} \biggl[ x^h \(
q^{-2h-4}y^h -(1+q^{-4}) y^{-h}\) \biggl. \nn \\
\biggr. && \hs{0.5cm}
 + x^{-h} \(q^{2h-8}y^{-h}-
q^{-2h+4}y^{3h}+ (q^{8}+q^{4}) y^h\)
-x^{-3h}q^{2h}y^h \biggr]\;.
\eea

Analogously, considering the right hand side of (\ref{BYBE2}), we see
that the first term is the $R$-matrix element $\cR_{i,i}^{k,l}(x)$,
which is zero for all but $k = l = i$. Therefore, the right hand side
is reduced to
\bea
\lefteqn{\sum_{\ol m,n} \cR_{i,i}^{i,i}(\frac xy)\, K_i^{\ol m}(x)\,
\cR_{i,\ol m}^{\ol  p,n}(xy)\, K_n^{\ol r}(y) =} \hs{0.5cm}
\nn \\
&=& \cR_{i,i}^{i,i}(\frac xy)\, \left[ K_i^{\ol i}(x)\,
\cR_{i,\ol i}^{\ol  p,p}(xy)\, K_p^{\ol r}(y) +
 K_i^{\ol p}(x)\,
\cR_{i,\ol p}^{\ol  p,i}(xy)\, K_i^{\ol r}(y) \right] \nn \\
&=&
( \(\frac xy \)^h - \(\frac xy \)^{-h} ) \(\frac xq
\)^{2i-2p} \(\frac yq \)^{2i-2r}
\left[ \(\frac qx\)^{2h} - \(\frac qy\)^{-2h} +x^{-h} (q^{4}+1)
( \(\frac yq\) ^{2h}  -q^{-4}) \right] \nn
\\
&& + (q^4-q^{-4}) \(\frac xq \)^{2i-2p} \(\frac yq \)^{2i-2r}
\left[ \(\frac{q^2}{xy} \)^h - \( \frac{q^2}{xy} \)^{-h} +
\(\frac yx \)^h (q^4+1) (1  - \(\frac qy\)^{2h} q^{-4}) \right] \nn \\
&=& \(\frac xq \)^{2i-2p} \(\frac yq \)^{2i-2r} \biggl[ x^h \(
q^{-2h-4}y^h -(1+q^{-4}) y^{-h}\) \biggr. \nn \\
&& \hs{0.5cm} \biggl. + x^{-h} \(q^{2h-8}y^{-h}-
q^{-2h+4}y^{3h}+ (q^{8}+q^{4}) y^h\)
-x^{-3h}q^{2h}y^h \biggr]\;,
\eea
which proves the BYBE for this particular case. We saw that we did not
need to know the explicit values of $i$ , $p$, $r$ or $n$. It was
sufficient to know which indices differ from each other. It is therefore
fairly straightforward to implement this in form of an algorithm and
check every case for general $n$. Here we have used MapleV to check
that (\ref{Ksol1}) and (\ref{Ksol2}) both solve (\ref{BYBE2}) for the
homogeneous as well as the principal gradation.
This procedure does not lend itself to finding new solutions easily,
but it is straightforward to check whether a given ansatz is a
solution.

\end{document}